\documentclass[cmbright,fleqn,referee]{envauth}

\def\bSig\mathbf{\Sigma}

\usepackage{natbib,upgreek}

\runninghead{M.B.~Hooten et al.}{Population-Level Inference for Animal Movement}

\begin{document}

\title{Hierarchical Animal Movement Models for Population-Level Inference}

\author{Mevin B. Hooten\affil{a}\corrauth\ Frances E. Buderman\affil{b}, Brian M. Brost\affil{b}, Ephraim M. Hanks\affil{c}, and Jacob S. Ivan\affil{d}}

\corraddr{M.B. Hooten, E-mail: mevin.hooten@colostate.edu}

\address{\affilnum{a}U.S. Geological Survey, Colorado Cooperative Fish and Wildlife Research Unit; Departments of Fish, Wildlife, \& Conservation Biology and Statistics, Colorado State University, Fort Collins, CO 80523\\
\affilnum{b}Department of Fish, Wildlife, and Conservation Biology, Colorado State University \\
\affilnum{c}Department of Statistics, Pennsylvania State University \\
\affilnum{d}Colorado Parks and Wildlife \\
}

\begin{abstract}
New methods for modeling animal movement based on telemetry data are developed regularly.  With advances in telemetry capabilities, animal movement models are becoming increasingly sophisticated.  Despite a need for population-level inference, animal movement models are still predominantly developed for individual-level inference.  Most efforts to upscale the inference to the population-level are either \emph{post hoc} or complicated enough that only the developer can implement the model.  Hierarchical Bayesian models provide an ideal platform for the development of population-level animal movement models but can be challenging to fit due to computational limitations or extensive tuning required.  We propose a two-stage procedure for fitting hierarchical animal movement models to telemetry data.  The two-stage approach is statistically rigorous and allows one to fit individual-level movement models separately, then resample them using a secondary MCMC algorithm.  The primary advantages of the two-stage approach are that the first stage is easily parallelizable and the second stage is completely unsupervised, allowing for a completely automated fitting procedure in many cases.  We demonstrate the two-stage procedure with two applications of animal movement models.  The first application involves a spatial point process approach to modeling telemetry data and the second involves a more complicated continuous-time discrete-space animal movement model.  We fit these models to simulated data and real telemetry data arising from a population of monitored Canada lynx in Colorado, USA.    
\end{abstract}

\maketitle

\section{Introduction}
The field of movement ecology is booming, in large part, because of the increased availability of telemetry data sources (\citealt{Cagnacci:10}).  Contemporary telemetry data are acquired via satellite communication devices affixed to individual animals.  These devices often collect many types of data, but most studies are focused on the position data, primarily to learn about environmental influences on individual-level movement.  Many new statistical models for animal trajectories have been proposed in recent years and they vary in form depending on the motivation for the project and type of inference desired (\citealt{Hooten:16}).  For example, most individual-based statistical models for telemetry data fall into one of three classes:  point process models, discrete-time models, or continuous-time models, with each being appropriate in certain settings (\citealt{McClintock:14}).  

Statistical inference arising from fitting animal movement models to telemetry data is sometimes focused on the individual level.  For example, a movement ecologist might ask how a specific individual animal responded to environmental cues while migrating between summer and winter home ranges (e.g., \citealt{Hooten:10a}).  However, many animal movement studies are concerned with population-level inference.  That is, for several individuals, is there evidence of consistent behavioral responses to environmental variables?  To obtain population-level inference, the well-accepted approach is to use a hierarchical model with random effects for individuals that are pooled at the population-level.  For example, consider the Bayesian hierarchical model  
\begin{align}
  \mathbf{y}_j &\sim [\mathbf{y}_j | \boldsymbol\beta_j,\boldsymbol\theta_j] \;, \label{eq:gen_datamodel} \\
  \boldsymbol\beta_j &\sim [\boldsymbol\beta_j | \boldsymbol\mu_\beta, \boldsymbol\Sigma_\beta] \;, \label{eq:gen_processmodel} \\
  \boldsymbol\mu_\beta &\sim [\boldsymbol\mu_\beta] \;, \label{eq:gen_mumodel} \\
  \boldsymbol\Sigma_\beta^{-1} &\sim [\boldsymbol\Sigma_\beta^{-1}] \;, \label{eq:gen_Sigmodel} \\
  \boldsymbol\theta_j &\sim [\boldsymbol\theta_j] \;, \label{eq:gen_auxparmodel}
\end{align}
\noindent where $\mathbf{y}_j$ are measurements associated with each individual $j$ ($j=1,\ldots,J$) and we use `$[\ldots]$' to denote a probability distribution or mass/density function as necessary (\citealt{GelfandSmith:90}).  The priors in (\ref{eq:gen_mumodel})--(\ref{eq:gen_auxparmodel}) are for the auxiliary data-level parameters $\boldsymbol\theta_j$, population-level coefficients $\boldsymbol\mu_\beta$, and precision matrix $\boldsymbol\Sigma_\beta^{-1}$, forming the familiar three-level hierarchical model (\citealt{Berliner:96}).  The hierarchical model in (\ref{eq:gen_datamodel})--(\ref{eq:gen_auxparmodel}) provides a straightforward and intuitive means for obtaining inference for $\boldsymbol\mu_\beta$, which is the ultimate goal of many animal movement studies.  Similar hierarchical models have become popular, and now standard, tools for obtaining upscaled inference in many other fields such as atmospheric science (\citealt{CressieWikle:11}), ecology (\citealt{HobbsHooten:15}), and sociology (\citealt{GelmanHill:06}). 

The complexity of modern animal movement models makes implementation challenging.  Furthermore, increases in the quantity of data resulting from newer telemetry devices has outpaced computational methods for fitting animal movement models.  Animal ecologists may wish to extend individual-level models to provide statistically rigorous population-level inference, but, in many cases, the algorithms required to fit such models become prohibitively challenging to program or are too slow in settings with large data sets and/or many individuals.  For example, \cite{Hanks:11} performed a \emph{post hoc} meta-analysis to obtain population-level inference for northern fur seals (\emph{Callorhinus ursinus}) because the implementation of a full hierarchical movement model was not computationally feasible.  Furthermore, in the Bayesian setting, Markov Chain Monte Carlo (MCMC) algorithms for most animal movement models require tuning from the user due to lack of conjugacy.  In cases where data sets from tens or hundreds of individuals are available, it may not be feasible to tune individual-level Metropolis-Hastings updates for all parameters. 

We present a statistically rigorous two-stage procedure for economizing hierarchical animal movement models to provide exact population-level inference using a sequence of algorithms that are fast, stable, and require little or no tuning by the user.  Our approach is simple.  First, we fit individual-level models (\ref{eq:gen_datamodel}) independently using a preferred stochastic sampling algorithm.  Independent model fits in the first stage allow for parallel processing, leading to an improvement in computational efficiency that scales with the number of processors.  Second, we obtain exact population-level inference using a secondary MCMC algorithm that requires no tuning.  The secondary algorithm is based on a little-known technique for Bayesian meta-analysis proposed by \cite{Lunn:13}.  We found that our two-stage procedure provides substantial computational improvements in both speed and ease of use in cases with large data sets and/or complicated data models.

In what follows, we present a general two-stage procedure for fitting a broad class of hierarchical animal movement models.  We then demonstrate the approach for a basic point process model for telemetry data (i.e., resource selection function model) and verify it using simulation.  In our second application, we show how the approach can be applied to a continuous-time discrete-space (CTDS) animal movement model using telemetry data with complicated error structure.  We apply the CTDS model to satellite telemetry data from a population of Canada lynx (\emph{Lynx canadensis}) in Colorado, USA.  Finally, we close with a summary and discussion of the approach and future directions.      

\section{Two-stage Procedure}
Many animal movement models have been constructed solely for individual-level inference (e.g., \citealt{Jonsen:05}; \citealt{Johnson:08b}; \citealt{Hooten:10a}; \citealt{Brost:15}; \citealt{Buderman:16}).  However, the desired scientific inference is usually at the population-level to assess if the population, as a whole, is responding to certain environmental cues.  Hierarchical statistical models provide a natural framework for obtaining upscaled population-level inference (\citealt{GelmanHill:06}; \citealt{HobbsHooten:15}).  As the complexity of the animal movement models increases, hierarchical models that include nonlinear components become challenging to implement due to computational limitations and user supervision requirements.  It is often much simpler to fit individual-level models to data, as long as individuals are assumed independent.  Following \cite{Lunn:13}, we propose a simple two-stage procedure for obtaining population-level inference under the full hierarchical model.  The two-stage procedure only requires independent individual-level model fits and an unsupervised resampling algorithm to obtain population-level inference without any user tuning. 

The first stage in the procedure involves fitting a data model like (\ref{eq:gen_datamodel}) independently for each individual $j$ ($j=1,\ldots,J$).  In addition to the prior for auxiliary data-level parameters $\boldsymbol\theta_j$ from (\ref{eq:gen_auxparmodel}), we also specify a prior for the individual-level parameters $\boldsymbol\beta_j$ as $\boldsymbol\beta_j \sim [\boldsymbol\beta_j]$ (where the priors for $\boldsymbol\theta_j$ and $\boldsymbol\beta_j$ can differ by individual).  The priors for $\boldsymbol\beta_j$ are only used in the first stage of the two-stage procedure and do not affect the final inference.  The posterior distribution for individual $j$ is           
\begin{equation} 
  [\boldsymbol\theta_j, \boldsymbol\beta_j | \mathbf{y}_j] = \frac{[\mathbf{y}_j | \boldsymbol\beta_j, \boldsymbol\theta_j][\boldsymbol\beta_j][\boldsymbol\theta_j]}{\int \int [\mathbf{y}_j | \boldsymbol\beta_j, \boldsymbol\theta_j][\boldsymbol\beta_j][\boldsymbol\theta_j] d\boldsymbol\beta_j d\boldsymbol\theta_j} \; . 
  \label{eq:first_post}
\end{equation} 
\noindent In principle, any stochastic sampling algorithm can be used to obtain samples from the posterior distribution in (\ref{eq:first_post}), but those relying on MCMC are most commonly applied in the animal movement literature.  However, because we treat the models in (\ref{eq:first_post}) for all $J$ individuals independently in the first stage, they can be fit in parallel using readily available software (e.g., the `parallel' R package;  \citealt{R:16}).  Additionally, if we choose a sampling algorithm for fitting the models in (\ref{eq:first_post}) that is unsupervised (i.e., requiring no supervised tuning), then the entire two-stage procedure can be automated.  An unsupervised fitting procedure will be used much more often by ecologists in situations where data exist for a large number of individuals.  Thus, automatic MCMC algorithms like BUGS (\citealt{Lunn:09}), JAGS (\citealt{Plummer:03}), or STAN (\citealt{Carpenter:16}) can be used to fit the individual-level models in (\ref{eq:first_post}), or alternatively, importance sampling or particle filtering (e.g., LibBi; \citealt{Murray:13}) methods can also be employed.  Finally, the choice of priors $[\boldsymbol\beta_j]$ can also lead to fully automatic and parallelizable first-stage algorithms.  For example, if the data model (\ref{eq:gen_datamodel}) is Poisson (i.e., $\mathbf{y}_j \sim \text{Pois}(\exp(\mathbf{X}_j\boldsymbol\beta_j))$, where $\mathbf{X}_j$ is a design matrix of covariates for the $j$th individual), then $\boldsymbol\theta_j$ is empty because the Poisson does not have a separate dispersion parameter.  A multivariate log-gamma prior distribution (\citealt{Crooks:10}; \citealt{Bradley:15}) for $\boldsymbol\beta_j$ facilitates the use of a Monte Carlo sampler to obtain posterior samples from (\ref{eq:first_post}).  For non-conjugate priors, adaptively tuned MCMC algorithms (e.g., \citealt{GivensHoeting:12}) are straightforward to implement and provide a way to obtain unsupervised stage-one samples for $\boldsymbol\beta_j$.        

The second stage in the two-stage procedure involves an MCMC algorithm resembling that used to fit the full hierarchical model, but with a critical simplification.  To fit the full hierarchical model in (\ref{eq:gen_datamodel})--(\ref{eq:gen_auxparmodel}), we sequentially sample from the full-conditional distributions $[\boldsymbol\beta_j | \cdot]$ for $j=1,\ldots,J$, $[\boldsymbol\mu_\beta|\cdot]$, and $[\boldsymbol\Sigma_\beta^{-1}|\cdot]$, using an MCMC algorithm.  In our second stage algorithm, we use the MCMC algorithm for the full hierarchical model as a template, but modify the updates for $\boldsymbol\beta_j$.  Updates for the individual-level auxiliary parameters, $\boldsymbol\theta_j$, are automatically coupled with those from $\boldsymbol\beta_j$, but are only necessary if we desire inference for $\boldsymbol\theta_j$.  In fact, if $\boldsymbol\theta_j$ are considered nuisance parameters, it is not necessary to store samples for them in our two-stage procedure.  

The full-conditional distributions for population-level parameters $\boldsymbol\mu_\beta$ and $\boldsymbol\Sigma_\beta^{-1}$ in the second stage model remain the same as in the MCMC algorithm to fit the full hierarchical model in (\ref{eq:gen_datamodel})--(\ref{eq:gen_auxparmodel}):
\begin{align}
  [\boldsymbol\mu_\beta|\cdot] &\propto \left(\prod_{j=1}^J [\boldsymbol\beta_j | \boldsymbol\mu_\beta, \boldsymbol\Sigma_\beta]\right)[\boldsymbol\mu_\beta] \;, \label{eq:mu_fullcond} \\
  [\boldsymbol\Sigma_\beta^{-1}|\cdot] &\propto \left(\prod_{j=1}^J [\boldsymbol\beta_j | \boldsymbol\mu_\beta, \boldsymbol\Sigma_\beta]\right)[\boldsymbol\Sigma_\beta^{-1}] \;. \label{eq:Sig_fullcond}
\end{align}
\noindent If the model for $\boldsymbol\beta_j$ and prior for $\boldsymbol\mu_\beta$ are multivariate Gaussian and the prior for $\boldsymbol\Sigma_\beta^{-1}$ is Wishart, then the full-conditional distributions in (\ref{eq:mu_fullcond}) and (\ref{eq:Sig_fullcond}) are multivariate Gaussian and Wishart, respectively.  These specific distributions are commonly used in many animal movement models for population-level parameters and permit conjugate Gibbs updates in our second stage algorithm.    

The joint full-conditional distribution for the data-level auxiliary parameters, $\boldsymbol\theta_j$, and individual-level parameters, $\boldsymbol\beta_j$, is  
\begin{equation} 
  [\boldsymbol\theta_j, \boldsymbol\beta_j |\cdot] \propto [\mathbf{y}_j | \boldsymbol\beta_j, \boldsymbol\theta_j][\boldsymbol\beta_j|\boldsymbol\mu_\beta, \boldsymbol\Sigma_\beta][\boldsymbol\theta_j] \;,  
  \label{eq:beta_fullcond}
\end{equation}
\noindent which, depending on the form of data model $[\mathbf{y}_j | \boldsymbol\beta_j,\boldsymbol\theta_j]$, would normally require a Metropolis-Hastings update.  In this case, the Metropolis-Hastings ratio for the joint update of $\boldsymbol\theta_j$ and $\boldsymbol\beta_j$ is   
\begin{equation} 
  r_j = \frac{[\mathbf{y}_j | \boldsymbol\beta_j^*,\boldsymbol\theta_j^*][\boldsymbol\beta_j^*|\boldsymbol\mu_\beta^k, \boldsymbol\Sigma_\beta^k][\boldsymbol\theta_j^*][\boldsymbol\theta_j^{k-1},\boldsymbol\beta_j^{k-1}|\boldsymbol\theta_j^*,\boldsymbol\beta_j^*]}{[\mathbf{y}_j | \boldsymbol\beta_j^{k-1},\boldsymbol\theta_j^{k-1}][\boldsymbol\beta_j^{k-1}|\boldsymbol\mu_\beta^k, \boldsymbol\Sigma_\beta^k][\boldsymbol\theta_j^{k-1}][\boldsymbol\theta_j^*,\boldsymbol\beta_j^*|\boldsymbol\theta_j^{k-1},\boldsymbol\beta_j^{k-1}]} \; , 
  \label{eq:mh_hier_beta}
\end{equation}
\noindent where, the `$*$' superscript represents the proposal for $\boldsymbol\beta_j$ and the `$k$' and `$k-1$' superscripts correspond to the MCMC sample for the $k$ or $k-1$ iteration of the MCMC algorithm (for $k=2,\ldots,K$). Typically, the proposal distribution, $[\boldsymbol\theta_j^*,\boldsymbol\beta_j^*|\boldsymbol\theta_j^{k-1},\boldsymbol\beta_j^{k-1}]$, is chosen to be a multivariate Gaussian random walk such that $(\boldsymbol\theta_j^*, \boldsymbol\beta_j^*)' \sim \text{N}((\boldsymbol\theta_j^{k-1}, \boldsymbol\beta_j^{k-1})',\tilde{\boldsymbol\Sigma}_j)$ which requires tuning for each individual $j$ by adjusting $\tilde{\boldsymbol\Sigma}_j$ using trial and error or an adaptive MCMC approach (e.g., \citealt{RobertsRosenthal:09}). 

However, if we use the posterior samples for $\boldsymbol\theta_j$ and $\boldsymbol\beta_j$ from the first stage (\ref{eq:first_post}) as the proposal in the second stage update for $\boldsymbol\beta_j$, then the proposal distribution is 
\begin{equation} 
  [\boldsymbol\theta_j^*,\boldsymbol\beta_j^*|\boldsymbol\theta_j^{k-1},\boldsymbol\beta_j^{k-1}] \equiv \frac{[\mathbf{y}_j | \boldsymbol\beta_j^*,\boldsymbol\theta_j^*][\boldsymbol\beta^*][\boldsymbol\theta_j^*]}{\int \int [\mathbf{y}_j | \boldsymbol\beta_j,\boldsymbol\theta_j][\boldsymbol\beta_j][\boldsymbol\theta_j] d\boldsymbol\beta_j d\boldsymbol\theta_j} \; , 
\end{equation} 
\noindent which does not depend on the previous $\boldsymbol\theta_j^{k-1}$ and $\boldsymbol\beta_j^{k-1}$.  The Metropolis-Hastings ratio from (\ref{eq:mh_hier_beta}) simplifies to
\begin{equation} 
  r_j = \frac{[\boldsymbol\beta_j^*|\boldsymbol\mu_\beta^k, \boldsymbol\Sigma_\beta^k][\boldsymbol\beta_j^{k-1}]}{[\boldsymbol\beta_j^{k-1}|\boldsymbol\mu_\beta^k, \boldsymbol\Sigma_\beta^k][\boldsymbol\beta_j^*]} \; , 
  \label{eq:mh_2_beta}
\end{equation}
\noindent while the updates for $\boldsymbol\mu_\beta$ and $\boldsymbol\Sigma_\beta^{-1}$ remain unchanged.  Thus, we keep the samples for $\boldsymbol\theta_j^*$ and $\boldsymbol\beta_j^*$, from the first stage, with probability $\text{min}(r_j,1)$.  However, we only need to explicitly save samples for the auxiliary individual-level parameters ($\boldsymbol\theta_j$) in the first or second stages if we desire inference on them because $r_j$, from (\ref{eq:mh_2_beta}), does not depend on $\boldsymbol\theta_j$.  Furthermore, \cite{Lunn:13} note that, when the stage one priors for $\boldsymbol\beta_j$ are diffuse, the ratio simplifies further to  $r_j = [\boldsymbol\beta_j^*|\boldsymbol\mu_\beta^k, \boldsymbol\Sigma_\beta^k]/[\boldsymbol\beta_j^{k-1}|\boldsymbol\mu_\beta^k, \boldsymbol\Sigma_\beta^k]$, a mere quotient involving the individual-level process distributions.  However, we retain the form in (\ref{eq:mh_2_beta}) so that we can use prior information when available.  Because there is no Markov dependence in the proposal for $\boldsymbol\beta_j$, we select $\boldsymbol\beta_j^*$ (and $\boldsymbol\theta_j^*$, if desired) uniformly at random from the output resulting from the first stage model fits.  More importantly, the Metropolis-Hastings ratios ($r_j$, for $j=1,\ldots,J$) in (\ref{eq:mh_2_beta}) do not contain a tuning parameter, resulting in unsupervised updates.  Paired with the Gibbs updates for $\boldsymbol\mu_\beta$ and $\boldsymbol\Sigma_\beta^{-1}$, the second stage algorithm is fully automatic, and samples from the full-conditional for $\boldsymbol\beta_j$ can be obtained in parallel (within the broader second stage MCMC algorithm) creating the potential for additional computational efficiency.  Critically, the Metropolis-Hastings ratio, $r_j$ in (\ref{eq:mh_2_beta}), is not a function of the data.  Therefore, complicated data models do not need to be reconsidered in the second stage algorithm.  The utility of the simple two-stage procedure is that it is intuitive, facilitates parallelization, and can result in algorithms that are fully automatic.   

In what follows, we provide two example applications where the two-stage procedure for obtaining population-level animal movement inference is valuable.  The first application involves a spatial point process modeling approach for telemetry data commonly referred to as ``resource selection function'' (RSF) analysis (e.g., \citealt{Manly:07}).  The second application involves a continuous-time discrete-space animal movement model proposed by \cite{Hooten:10a} and \cite{Hanks:15a}.      

\section{Applications}
\subsection{Hierarchical Point Process Models}
Perhaps the most common model fit to temporally independent telemetry data is the RSF model.  The RSF model is a heterogeneous point process model that conditions on the number of telemetry observations.  Assuming there is no measurement error associated with the telemetry data $\mathbf{s}_{ij}$ (typically a $2\times 1$ vector) for observations $i=1,\ldots,n_j$ and individuals $j=1,\ldots,J$, the data model takes the form of a weighted distribution (\citealt{PatilRao:77}) such that $\mathbf{s}_{ij} \sim [\mathbf{s}_{ij} | \boldsymbol\beta_j]$ and  
\begin{equation} 
  [\mathbf{s}_{ij} | \boldsymbol\beta_j] \equiv \frac{g(\mathbf{x}(\mathbf{s}_{ij}),\boldsymbol\beta_j) f(\mathbf{s}_{ij})}{\int g(\mathbf{x}(\mathbf{s}),\boldsymbol\beta_j)f(\mathbf{s})d\mathbf{s}} \;,
  \label{eq:rsf_datamodel}
\end{equation}
\noindent where, $g(\mathbf{x}(\mathbf{s}),\boldsymbol\beta_j)$ is the ``selection'' function and $f(\mathbf{s})$ is the ``availability'' function.  Thus, the animal movement interpretation of (\ref{eq:rsf_datamodel}) is that inference for $\boldsymbol\beta_j$ provides insight about how individual $j$ selects resources (i.e., covariates, $\mathbf{x}$) from those available to it.  The selection function is often chosen to be exponential (i.e., $g(\mathbf{x}(\mathbf{s}_{ij}),\boldsymbol\beta_j)\equiv \exp(\mathbf{x}(\mathbf{s}_{ij})'\boldsymbol\beta_j)$) and the availability function is typically assumed to be uniform on the support of the point process (i.e., $f(\mathbf{s}_{ij})\equiv \text{unif}({\cal S})$ for $\mathbf{s}_{ij}\in {\cal S}\subset \Re\times\Re$).     

\cite{WartonShepherd:10} and \cite{Aarts:12} showed that the RSF model in (\ref{eq:rsf_datamodel}) can be fit using a variety of approaches, including a Poisson likelihood.  The Poisson likelihood can be considered by first preprocessing the data such that $\mathbf{y}_j\equiv (y_{1,j},\ldots,y_{m,j})'$ represents counts of telemetry locations in grid cells corresponding to a discretization of the support ${\cal S}$.  As the grid cell size decreases with respect to the resolution of the covariates $\mathbf{x}$, a Poisson data model coincides with the point process model.  Thus, the corresponding hierarchical model     
\begin{align}
  \mathbf{y}_j &\sim \text{Pois}(\exp(\mathbf{X}_j\boldsymbol\beta_j)) \;, \label{eq:spp1} \\
  \boldsymbol\beta_j &\sim \text{N}(\boldsymbol\mu_\beta,\boldsymbol\Sigma_\beta) \;, \label{eq:spp2} \\
  \boldsymbol\mu_\beta &\sim \text{N}(\boldsymbol\mu_0,\boldsymbol\Sigma_0) \;, \label{eq:spp3} \\
  \boldsymbol\Sigma_\beta^{-1} &\sim \text{Wish}((\mathbf{S}\nu)^{-1},\nu) \;, \label{eq:spp4} 
\end{align}
\noindent assumes the same form as (\ref{eq:gen_datamodel})--(\ref{eq:gen_auxparmodel}) and allows for population-level resource selection inference on $\boldsymbol\mu_\beta$.  To fit the full hierarchical model directly using MCMC, we sample from the full-conditional distributions for $\boldsymbol\beta_j$, $\boldsymbol\mu_\beta$, and $\boldsymbol\Sigma_\beta^{-1}$, sequentially.  Standard Metropolis-Hastings updates for $\boldsymbol\beta_j$ require tuning, but the model can be fit using a single MCMC algorithm for moderately sized data sets.  Alternatively, the weighted least squares proposal approach of \cite{Gamerman:97} could be used to acquire samples for $\boldsymbol\beta_j$ from the posterior distribution.  However, to adequately approximate the point process model, the grid cells often need to be quite small, resulting in a fine-scale discretization of the support ${\cal S}$ and increasing the computational burden.  

The two-stage procedure we described in the previous Section can easily be employed to fit the hierarchical model in (\ref{eq:spp1})--(\ref{eq:spp4}).  For the first stage, we can use an MCMC or Hamiltonian Monte Carlo algorithm (via BUGS, JAGS, or STAN; \citealt{Lunn:09}; \citealt{Plummer:03}; \citealt{Carpenter:16}) to fit the individual level models in parallel.  For our spatial point process setting, the individual-level models are 
\begin{align}
  \mathbf{y}_j &\sim \text{Pois}(\exp(\mathbf{X}_j\boldsymbol\beta_j)) \;, \label{eq:MSspp1} \\
  \boldsymbol\beta_j &\sim \text{N}(\boldsymbol\mu_0,\boldsymbol\Sigma_0) \;, \label{eq:MSspp2} 
\end{align}
\noindent for $j=1,\ldots,J$, independently.  Note that the individual-level parameter model in (\ref{eq:MSspp2}) is an exchangeable prior for all $j=1,\ldots,J$.  Also, if the individual data sets $\mathbf{y}_j$ and $\mathbf{X}_j$ are so large that they are difficult to store in memory simultaneously for all $J$ individuals, the first stage model fitting can be fully distributed among separate machines or performed in sequence.  This highlights another primary advantage of the two-stage procedure.    

The second stage algorithm for obtaining population-level inference is an MCMC algorithm with Gibbs updates for $\boldsymbol\mu_\beta$ and $\boldsymbol\Sigma_\beta^{-1}$ as described in the previous Section, and updates for $\boldsymbol\beta_j$ using Metropolis-Hastings based on the acceptance ratio in (\ref{eq:mh_2_beta}), which becomes  
\begin{equation} 
  r_j = \frac{\text{N}(\boldsymbol\beta_j^*|\boldsymbol\mu_\beta^k, \boldsymbol\Sigma_\beta^k)\text{N}(\boldsymbol\beta_j^{k-1}|\boldsymbol\mu_0,\boldsymbol\Sigma_0)}{\text{N}(\boldsymbol\beta_j^{k-1}|\boldsymbol\mu_\beta^k, \boldsymbol\Sigma_\beta^k)\text{N}(\boldsymbol\beta_j^*|\boldsymbol\mu_0,\boldsymbol\Sigma_0)} \; . 
  \label{eq:mh_app1}
\end{equation}

Within the second stage MCMC algorithm, the updates for $\boldsymbol\beta_j$ can also be parallelized because they are independent, although this model is simple enough that parallelization is not necessary in the second stage algorithm.  Thus, the data, $\mathbf{y}_j$ for $j=1,\ldots,J$, which could include counts for 10s or 100s of thousands of grid cells and 100s of individuals, do not appear in the second stage algorithm.  The absence of $\mathbf{y}_j$ leads to a more computationally efficient second stage algorithm than the original algorithm to fit the full hierarchical model directly.    

\begin{figure}[htp]
  \centering
  \includegraphics[width=5in, angle=0]{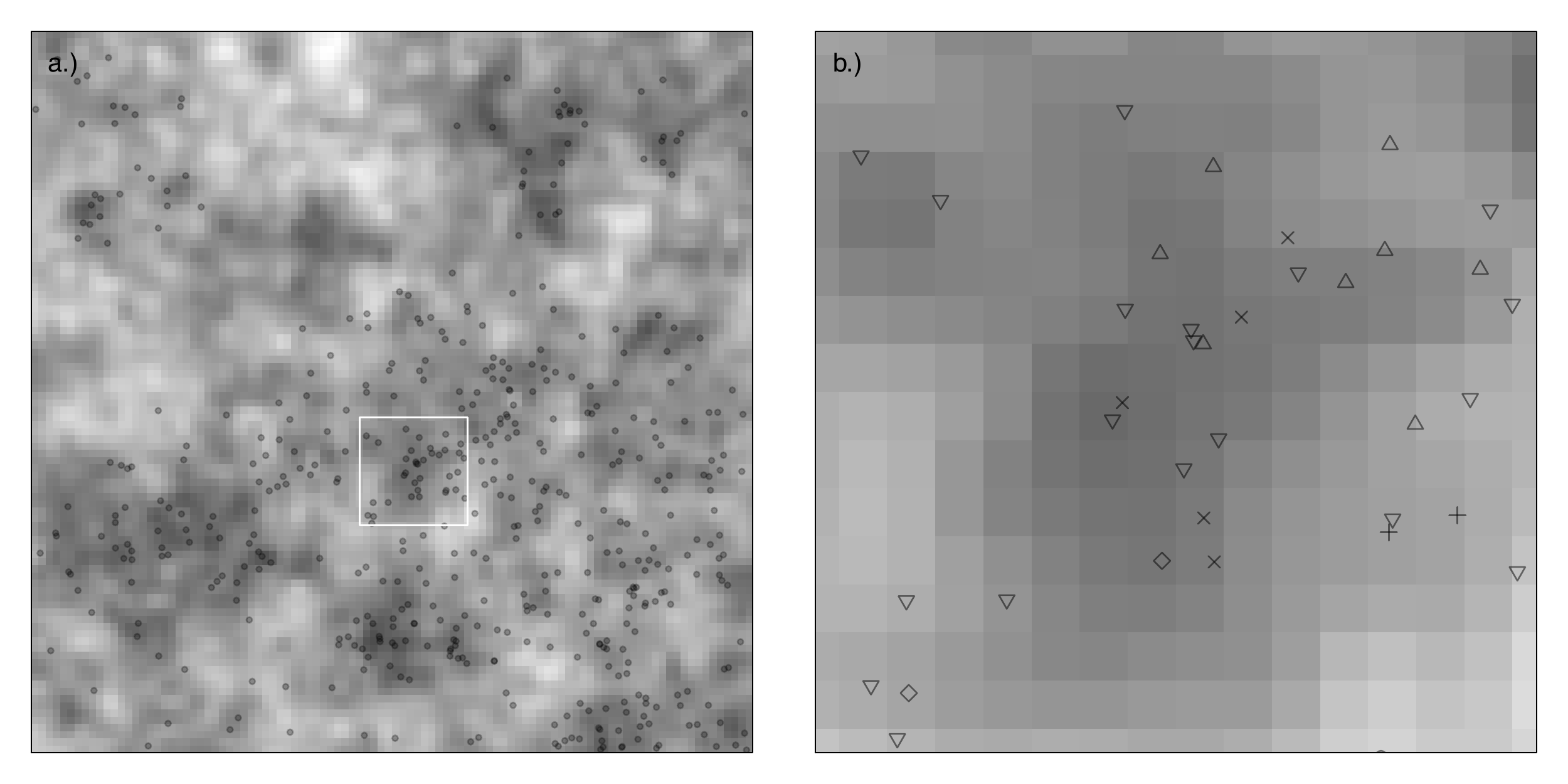}
  \caption{a.) Simulated animal positions (points) based on a spatial point process (\ref{eq:rsf_datamodel}) with one simulated covariate (background image, dark shading represents larger values).  b.) A zoomed in spatial map (from inset white box in panel a) showing positions from 5 individual animals as different point types (i.e., $\diamond$, $\bigtriangleup$, $\bigtriangledown$, $+$, $\times$).}
  \label{fig:rsf_data}
\end{figure}

We simulated point process data from 20 individuals (Figure~\ref{fig:rsf_data}), resulting in approximately 30 simulated telemetry fixes per individual, and fit the hierarchical RSF model using: 1.) a single MCMC algorithm, and 2.) our two-stage procedure.  We compared the population-level results from the fits resulting from each procedure.  
For the first-stage algorithm in our two-stage procedure, we fit the individual-level models independently using an adaptive MCMC algorithm in parallel using R (\citealt{R:16}) and assumed $\text{N}(\mathbf{0},100\cdot\mathbf{I})$ priors for $\boldsymbol\beta_j$, a $\text{N}(\mathbf{0},100\cdot\mathbf{I})$ prior for $\boldsymbol\mu_\beta$, and a $\text{Wish}((3\cdot\mathbf{I})^{-1},3)$ prior for $\boldsymbol\Sigma_\beta^{-1}$.  Our first-stage algorithm uses a multivariate Gaussian proposal for $\boldsymbol\beta_j$ and adapts the tuning using a single variance parameter, resulting in an unsupervised algorithm for the individual-level model fits.  We could have also used BUGS or JAGS to fit the first-stage models, but our adaptive MCMC algorithm required less computing time.  

The single MCMC algorithm to fit the full hierarchical model required 2.62 minutes to obtain 20,000 MCMC samples in R, whereas the first-stage algorithm required 0.57 minutes to obtain the same number of samples using an adaptive MCMC algorithm in parallel for the 20 individuals.  The second-stage algorithm required only 1.49 minutes in R, which implies that the total compute time to fit the model using the two-stage procedure was 2.06 minutes (0.56 minutes less than the single MCMC algorithm).  Also, the two-stage procedure requires no tuning and results in much larger effective MCMC sample sizes for parameters.  The effective MCMC sample sizes for $\boldsymbol\mu_\beta$ and $\boldsymbol\beta_j$ were 8560 and 1398 (averaged across individuals) for the single MCMC algorithm, but were 17590 and 15184 for the two-stage algorithm (out of 20,000 total samples).  Thus, to obtain the same effective MCMC sample size using MCMC for all parameters, we would need an order of magnitude more samples from the single MCMC algorithm.  

\begin{figure}[htp]
  \centering
  \includegraphics[height=7in, angle=0]{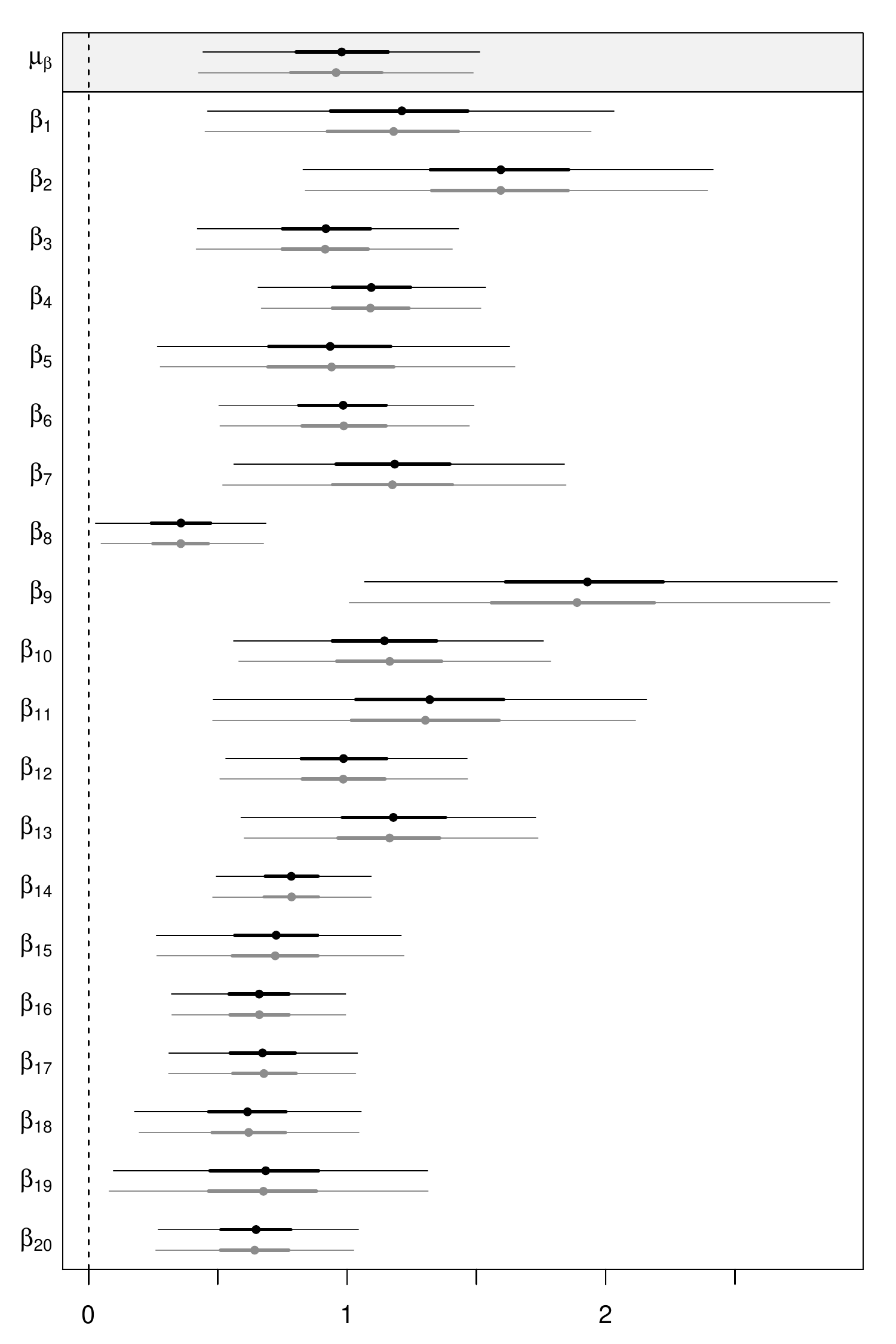}
  \caption{Posterior means (points) and 50\% and 95\% credible intervals for $\mu_{\beta 1}$ and $\beta_{j1}$ for $j=1,\ldots,20$.  Single MCMC algorithm results are shown in black and two-stage procedure results are shown in gray.}
  \label{fig:rsf_betas}
\end{figure}

Figure~\ref{fig:rsf_betas} illustrates the similarities in inference for the slope parameters $\mu_{\beta 1}$ and $\beta_{j1}$ for $j=1,\ldots,20$ when fitting the hierarchical RSF model using a single MCMC algorithm (black) versus the two-stage procedure (gray).    
Notice that the single MCMC algorithm and the two-stage procedure provide very similar inference.  In terms of inference, there exists some variability among individuals, but the population-level inference (Figure~\ref{fig:rsf_betas}, top) suggests a consistent overall positive population response to the covariate.     

\subsection{Hierarchical Continuous-Time Discrete-Space Models}
The previous application, involving spatial point process models, involves a commonly used model specification and desired type of inference in ecological research, but more contemporary methods have been developed to explicitly model the dynamics of animal movement based on temporally dependent telemetry data with observations close in time.  Among these methods are discrete-time and continuous-time approaches to modeling the individual animal trajectories (\citealt{McClintock:14}).  We focus on the continuous-time class of models in what follows.  

Continuous-time statistical models for animal movement processes have existed for decades (e.g., \citealt{DunnGipson:77}; \citealt{Blackwell:97}), and are usually based on Brownian motion (i.e., Wiener processes).  Up until the late 1990s, most Brownian motion models for trajectories utilized an Ornstein-Uhlenbeck process (i.e., a Wiener process with attraction to a central position).  \citealt{Johnson:08b} also proposed an Ornstein-Uhlenbeck model, but for the velocity (i.e., temporally differentiated position) rather than the position process.  \cite{HootenJohnson:16a}  generalized the continuous-time velocity models of \cite{Johnson:08b} in the context of Gaussian processes with covariance structure induced by temporal basis functions.  \cite{Buderman:16} used a simplified basis function parameterization to model Canada lynx (\emph{Lynx canadensis}) movement while accounting for measurement error in the telemetry data.  \cite{Buderman:16} refer to their model as a ``functional movement model'' and use it to provide inference for the true underlying continuous position process (i.e., $\boldsymbol\mu(t)$, for time $t$) of an individual. 

The approach developed by \cite{Buderman:16} assumes that the telemetry data $\mathbf{s}_{ij}$ are observed with error.  In fact, for the Canada lynx in our study, the bivariate measurement error follows an unusual X-shaped pattern because the telemetry data are collected by Service Argos (\citealt{Costa:10}) which relies on polar orbiting satellites.  Thus, \cite{Brost:15} and \cite{Buderman:16} developed a measurement error model based on a mixture distribution to account for the X-shaped Argos pattern (see Appendix A for details).  Properly accounting for measurement error adds another level to the hierarchical model in (\ref{eq:gen_datamodel})--(\ref{eq:gen_auxparmodel}) such that   
\begin{align}
  \mathbf{s}_{ij} &\sim [\mathbf{s}_{ij} | \boldsymbol\mu_j(t_i), \boldsymbol\phi_j] \;, \label{eq:gen_errormodel2} \\
  \mathbf{y}_j &\sim [\mathbf{y}_j | \boldsymbol\beta_j,\boldsymbol\theta_j] \;, \label{eq:gen_datamodel2} \\
  \boldsymbol\beta_j &\sim [\boldsymbol\beta_j | \boldsymbol\mu_\beta, \boldsymbol\Sigma_\beta] \;, \label{eq:gen_processmodel2} \\
  \boldsymbol\mu_\beta &\sim [\boldsymbol\mu_\beta] \;, \label{eq:gen_mumodel2} \\
  \boldsymbol\Sigma_\beta^{-1} &\sim [\boldsymbol\Sigma_\beta^{-1}] \;, \label{eq:gen_Sigmodel2} \\
  \boldsymbol\theta_j &\sim [\boldsymbol\theta_j] \;, \label{eq:gen_auxparmodel2} \\
  \boldsymbol\phi_j &\sim [\boldsymbol\phi_j] \;, \label{eq:gen_errorparmodel2}
\end{align}
\noindent for $j=1,\ldots,J$ individuals, and where $\mathbf{y}_j$ is an $m_j \times 1$ vector that represents a latent process that is linked to the true continuous position process $\{\boldsymbol\mu_j(t), \forall t\}$ by a deterministic functional $h$ such that $\mathbf{y}_j=h(\{\boldsymbol\mu_j(t), \forall t\})$, and $\boldsymbol\phi_j$ are measurement error covariance parameters.  

\cite{Hooten:10a} developed an individual-level animal movement model based on (\ref{eq:gen_errormodel2}) and (\ref{eq:gen_datamodel2}) where the latent variables $\mathbf{y}_j$ represent a sequential multinomial process indicating transitions among grid cells on a discretization of the spatial support ${\cal S}$.  The latent process model in (\ref{eq:gen_datamodel2}) relies on a continuous-time discrete-space (CTDS) representation of the position process.  However, because the functional $h(\cdot)$, that links the position process with the data, is non-invertible in their model, \cite{Hooten:10a} proposed a Bayesian multiple imputation procedure to account for uncertainty in the true position process when making inference on $\boldsymbol\beta_j$.  The multiple imputation procedure used by \cite{Hooten:10a} differs from the two-stage procedure we described herein because it does not allow for feedback from the individual-level parameters $\boldsymbol\beta_j$ to the position process $\{\boldsymbol\mu_j(t), \forall t\}$ or measurement error parameters $\boldsymbol\phi_j$.  \cite{Hooten:10a} used an imputation model to interpolate the position process and then integrated over the uncertainty in the position process while fitting (\ref{eq:gen_datamodel2}) to provide posterior inference for the individual-level parameters $\boldsymbol\beta_j$.  

\cite{Hanks:15a} showed that the multinomial process of \cite{Hooten:10a} could be reparameterized such that $[\mathbf{y}_j | \boldsymbol\beta_j,\boldsymbol\theta_j]$ can be modeled using Poisson regression.  Specifically, let $\tau_{cj}$ represent the amount of time individual $j$ remains in a grid cell for the $c$th ``stay/move'' pair associated with the discretization of the individual's path through a landscape (for $c=1,\ldots,n_j$).  Then let $y_{clj} \sim \text{Pois}(\tau_{cj}\exp(\mathbf{x}'_{clj}\boldsymbol\beta_j))$ where the index $l=1,2,4,5$ ($l=3$ is not necessary because corresponds to the middle cell which is captured by $\tau_{cj}$) denotes moves to neighboring grid cells in each cardinal direction (i.e., north, east, south, west).  That is, if individual $j$  moved north for ``stay/move'' pair $c$, then the data point $y_{c1j}=1$ and $y_{c2j}=y_{c3j}=y_{c4j}=0$ (see Appendix B for details).  The Poisson reparametrization dramatically improves computational efficiency at the individual level because the total number of observations used in the model ($4 m_j$) is a function of the grid cell size rather than the position process discretization as used in \cite{Hooten:10a}.  Thus, \cite{Hanks:15a} were able to fit the CTDS model to large telemetry data sets in a fraction of the time required by the multinomial method developed by \cite{Hooten:10a}.  However, neither \cite{Hooten:10a} nor \cite{Hanks:15a} attempted to fit a hierarchical model like that in (\ref{eq:gen_errormodel2})---(\ref{eq:gen_errorparmodel2}) to obtain population level inference for $\boldsymbol\mu_\beta$.         

In our application involving population-level inference for Canada lynx, we use the model developed by \cite{Buderman:16} to obtain the imputation distribution for the true individual-level position process $\{\tilde{\boldsymbol\mu}_j(t), \forall t\}$, and hence $\tilde{y}_{clj}$ for all $c$, $l$, and $j$, while accounting for the complicated nature of Argos telemetry error (see Appendix A for details).  In what follows, we combine all $\tilde{y}_{clj}$ into a single vector representing the latent process $\tilde{\mathbf{y}}_j$ and use $\tilde{\mathbf{y}}_j$ as data in a two-stage implementation of the hierarchical model in (\ref{eq:gen_errormodel2})--(\ref{eq:gen_errorparmodel2}).     

To fit the hierarchical model using the two-stage procedure described in Section 2, we apply the same two stages of algorithms as in the previous application.  For the first stage, we use the data model in (\ref{eq:gen_datamodel2}) and specify multivariate Gaussian priors for the individual-level parameters $\boldsymbol\beta_j \sim \text{N}(\boldsymbol\mu_0,\boldsymbol\Sigma_0)$.  We use an adaptively-tuned MCMC algorithm to obtain samples from the posterior distributions 
\begin{equation}
  [\boldsymbol\beta_j | \{\mathbf{s}_{ij}, \forall i, j\}]= \int [\boldsymbol\beta_j | \tilde{\mathbf{y}}_j][\tilde{\mathbf{y}}_j | \{\mathbf{s}_{ij}, \forall i, j\}] d\tilde{\mathbf{y}}_j \; ,
  \label{eq:mult_imp}
\end{equation}
\noindent for $j=1,\ldots,J$, and where, $[\tilde{\mathbf{y}}_j | \{\mathbf{s}_{ij}, \forall i, j\}]$ represents the imputation distribution for the latent Poisson process.  To perform the integration in (\ref{eq:mult_imp}), we simply sample $\tilde{\mathbf{y}}^k_j \sim [\tilde{\mathbf{y}}_j | \{\mathbf{s}_{ij}, \forall i, j\}]$ on the $k$th MCMC iteration and then let the Metropolis-Hastings update $\boldsymbol\beta^k_j$ depend on $\tilde{\mathbf{y}}^k_j$ as described in \cite{Hooten:10a} and \cite{Hanks:15a}.  As in the first application, we can fit the $J$ models for all individuals in parallel, dramatically reducing the required computational time.  

For the second stage of the two-stage procedure, we use the posterior samples for $\{\boldsymbol\beta_j, \forall j\}$, from the first stage, as proposals in the MCMC algorithm to fit the hierarchical model in (\ref{eq:gen_datamodel2})--(\ref{eq:gen_Sigmodel2}).  In doing so, we update $\{\boldsymbol\beta_j, \forall j\}$, $\boldsymbol\mu_\beta$, and $\boldsymbol\Sigma_\beta^{-1}$ sequentially in a completely unsupervised second-stage MCMC algorithm.  Recall that the Metropolis-Hastings acceptance ratio for $\boldsymbol\beta_j$ is identical to that used in the previous application (\ref{eq:mh_app1}).  As a result of the two-stage implementation and the adaptive tuning in the first-stage algorithm, the procedure is completely automatic after the data are preprocessed to obtain the imputation distribution, and population-level inference for $\boldsymbol\mu_\beta$ can easily be obtained.        

\begin{figure}[htp]
  \centering
  \includegraphics[height=6in, angle=0]{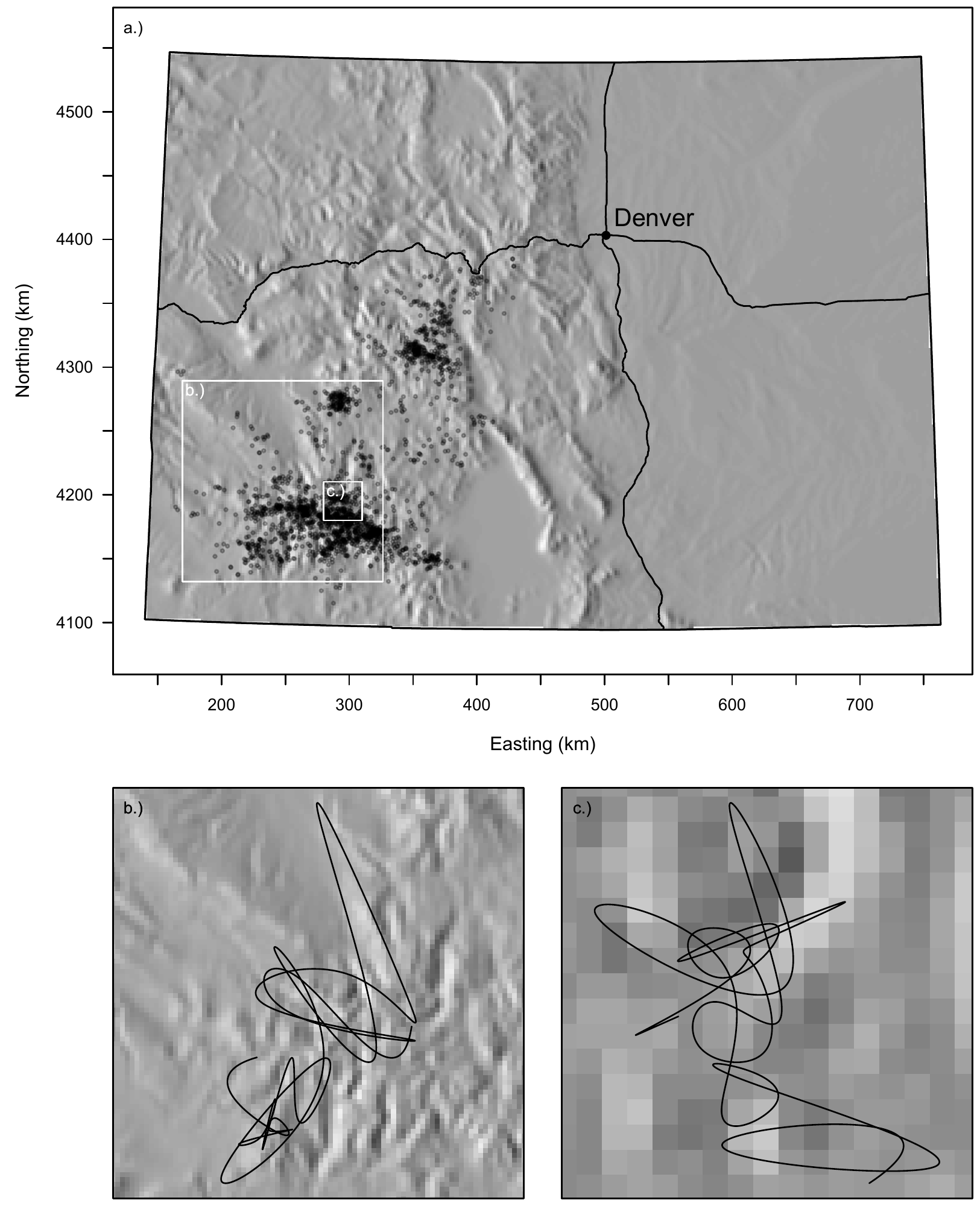}
  \caption{a.) Colorado, USA, with major highways and the city of Denver shown.  The telemetry data spanning a year of time for 18 individual Canada lynx are shown as points.  A shaded relief map is shown as the background image to illustrate the topography of the area.  b.) and c.) Close up views of the predicted paths for two individual Canada lynx.  For clarity, only the posterior mean path is shown.}
  \label{fig:lynx_data}
\end{figure}

\begin{figure}[htp]
  \centering
  \includegraphics[width=5in, angle=0]{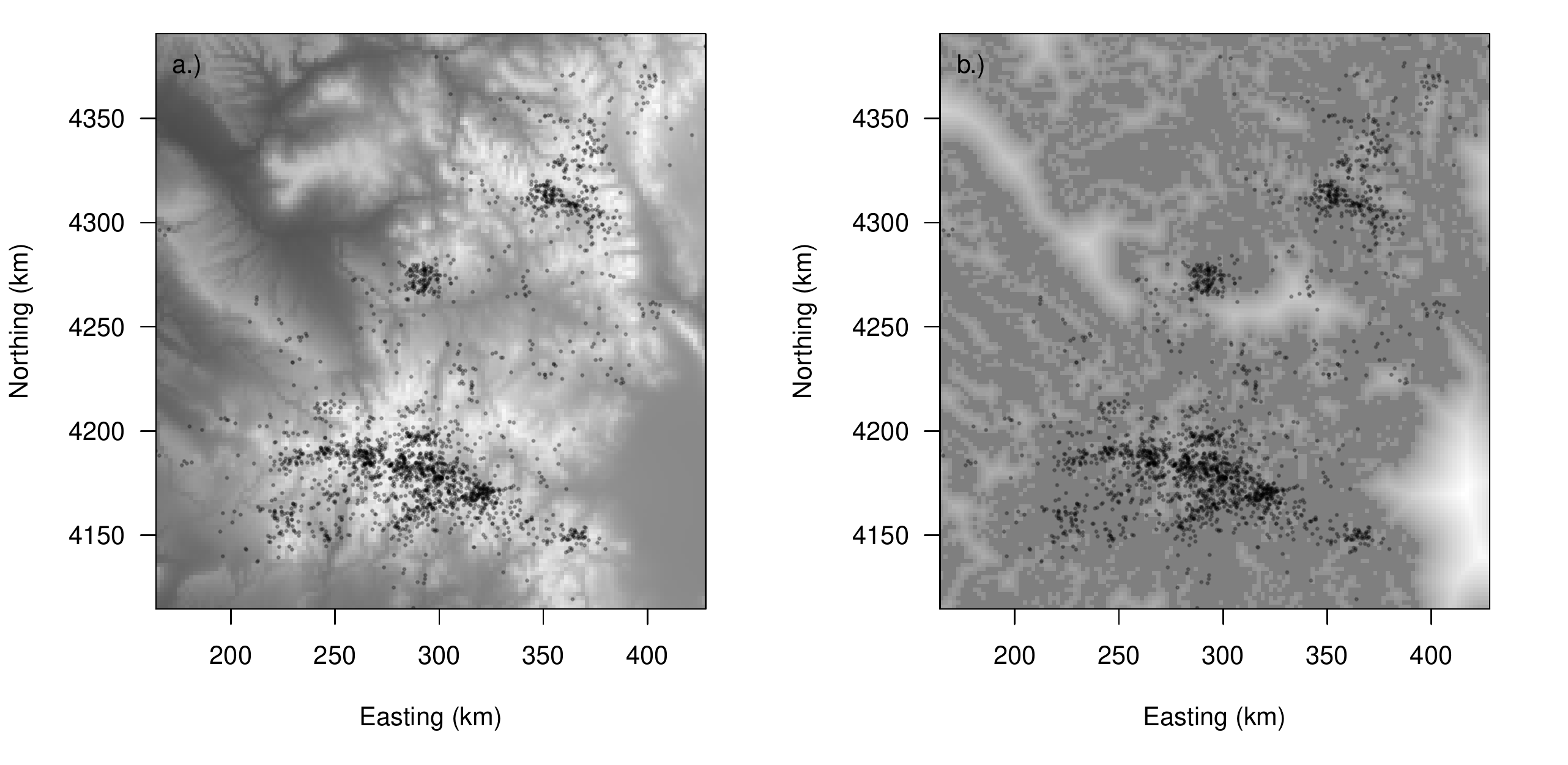}
  \caption{Images of covariates with telemetry observations overlaid as black points: a.) Elevation and b.) distance to forest.  Light shading corresponds to larger values.}
  \label{fig:lynx_covs}
\end{figure}

Using telemetry data from $J=18$ individual Canada lynx in Colorado, USA (Figure~\ref{fig:lynx_data}a), we applied the two-stage procedure to fit the hierarchical model in (\ref{eq:gen_datamodel2})--(\ref{eq:gen_Sigmodel2}).  
We used the functional movement model of \cite{Buderman:16} to obtain the imputed path distribution (Figure~\ref{fig:lynx_data}b,c) for each individual and used nearly continuous imputed path realizations to create the latent Poisson data realizations $\tilde{\mathbf{y}}^k_j$ (resulting in approximately 450 discrete-space transitions per individual, $n_j\approx 450$).  Canada lynx are a subalpine species that tend to prefer forested ecosystems (\citealt{McKelvey:00}), thus we focused on two covariates: elevation and distance to forest (Figure~\ref{fig:lynx_covs}).  
Each covariate was included in the model as a ``static'' driver, rather than a gradient-based driver of movement (\citealt{Hanks:15a}).  Static drivers can be interpreted as affecting overall motility in the CTDS model.  For priors in the first stage, we used $\boldsymbol\beta_j\sim \text{N}(\mathbf{0},100\mathbf{I})$ for all $j=1,\ldots,18$.  We used $\boldsymbol\mu_\beta \sim \text{N}(\mathbf{0},100\mathbf{I})$ and $\boldsymbol\Sigma_\beta^{-1} \sim \text{Wish}((3\cdot\mathbf{I})^{-1},3)$ as priors for the population-level parameters and precision matrix.  See Appendix B for additional details on the CTDS animal movement model.  

\begin{figure}[htp]
  \centering
  \includegraphics[width=5in, angle=0]{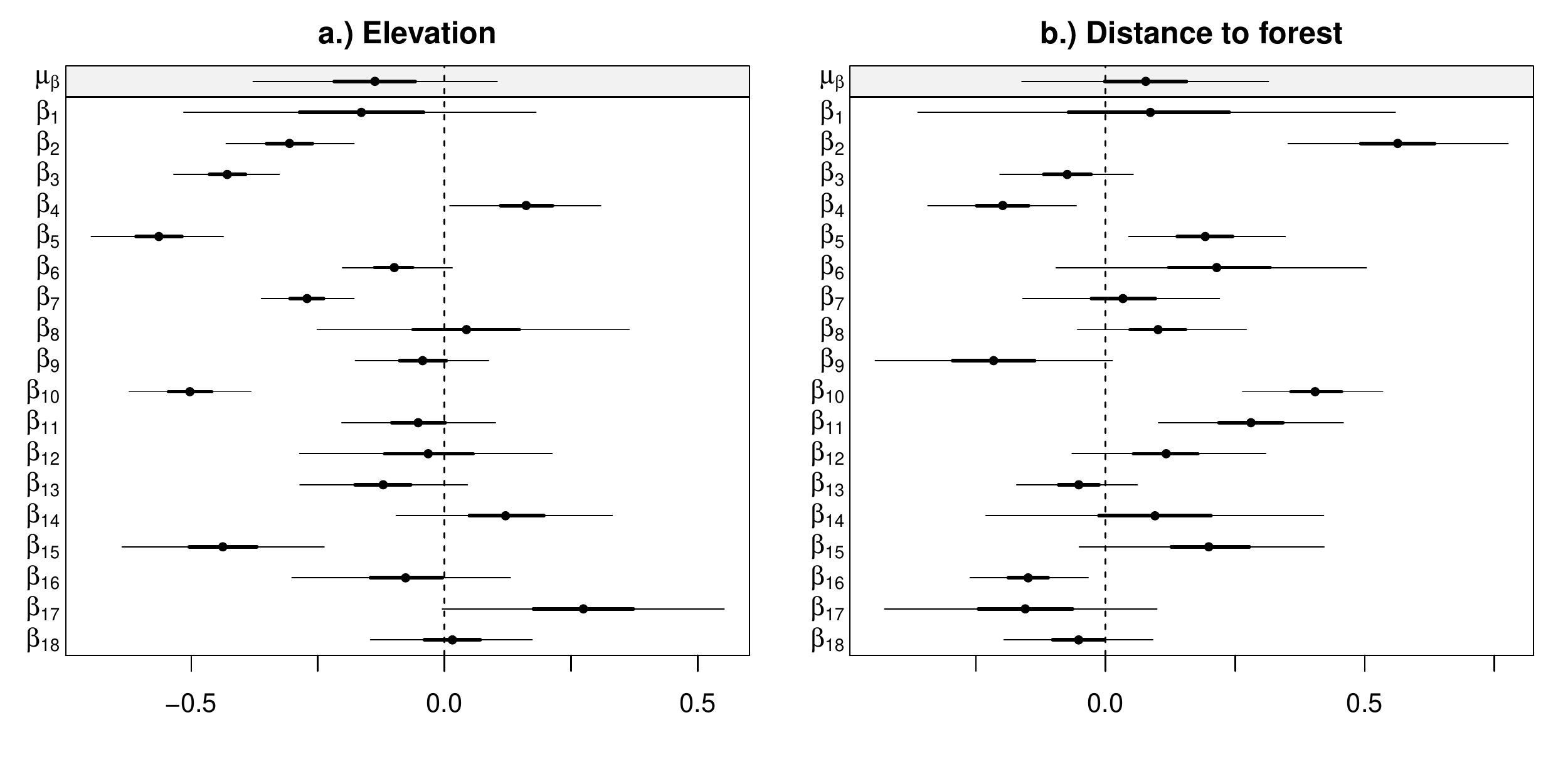}
  \caption{Posterior estimates for the population-level parameters $\boldsymbol\mu_\beta$ and individual-level parameters $\boldsymbol\beta_j$.  The posterior mean is shown as a central point and the 50\% and 95\% credible intervals are shown as the thick and thin black lines.  Panel a shows the results for the elevation covariate and panel b shows the results for the distance from forest covariate.}
  \label{fig:lynx_betas}
\end{figure}

We fit the overall hierarchical model using the two-stage procedure and the resulting algorithms required 0.86 minutes for the first stage (using an adaptive MCMC algorithm in parallel) and 1.62 minutes for the second stage.  Figure~\ref{fig:lynx_betas} shows the results of the model fit in terms of posterior means and 50\% and 95\% credible intervals for the population-level parameters $\boldsymbol\mu_\beta$ and individual-level parameters $\boldsymbol\beta_j$.  
While there exists substantial variability among individual Canada lynx, with some individuals exhibiting clear relationships with the covariates (e.g., individuals 2, 4, and 5), the posterior distributions for $\boldsymbol\mu$ did not indicate a population-level effect for either covariate at the 95\% level (but both did at the 50\% level).  For the individuals that did show evidence of an effect (i.e., 95\% credible intervals not overlapping zero), the negative response to elevation indicates that overall motility decreases at higher elevations, leading to greater residence times in those regions, as opposed to lower elevations (Figure~\ref{fig:lynx_betas}a).  Similarly, for individuals with significant effects related to distance from forest we see positive influence on motility implying that those Canada lynx have higher motility (and hence lower residence time) in regions farther from forest (Figure~\ref{fig:lynx_betas}b).  Thus, the inference in our application involving Canada lynx agrees with that obtained in other studies (e.g., \citealt{McKelvey:00}).     

\section{Conclusion}
Our findings indicate that the two-stage procedure we described herein holds tremendous value for fitting hierarchical animal movement models to telemetry data for population-level inference.  We applied the two-stage procedure to two types of commonly used animal movement models of varying complexity and found that it worked well in both cases.  

The spatial point process modeling approach we described in the first application is a commonly used model, but still fairly simple.  Much more complicated spatio-temporal point process models have been used to model temporally correlated telemetry data (e.g., \citealt{Johnson:08a}; \citealt{Johnson:13}; \citealt{Brost:15}) and adapting the two-stage procedure to those models is the subject of ongoing research.  For example, \cite{Brost:15} developed a model with a time-varying dynamic availability component that depended on an additional smoothness parameter.  Thus, the data model developed by \cite{Brost:15} required substantially more computation time than the simulated example we presented in Section 3.1 and would benefit from a two-stage implementation where individual-level models could be fit independently on separate processors and then recombined using the second stage MCMC algorithm to yield population-level inference for $\boldsymbol\mu_\beta$.        

In our example involving Canada lynx, the continuous-time discrete-space reparameterization developed by \cite{Hanks:15a} already provides significant improvements in computational efficiency over the motivating model developed by \cite{Hooten:10a}.  However, additional computational gains can be achieved using the two-stage fitting procedure to provide population-level inference.  

Despite the wide range of potential applications to many types of hierarchical models, we found it surprising that the two-stage fitting procedure of \cite{Lunn:13} is not more well known.  For our situations with large amounts of telemetry data and potentially complicated data models, we found the two-stage procedure works very well and is trivial to implement.  We also found it very helpful to be able to use different data models, first-stage fitting algorithms, and easy parallelization.  As a potential caveat, the two-stage procedure described by \cite{Lunn:13} may not be very efficient when the population induces extreme amounts of shrinkage in the individual-level parameters. Thus, in these cases, more samples would be needed in the first stage algorithm.  However, in a preliminary simulation study, we found that the two-stage procedure performs poorly only for data sets with very small amounts of data (i.e., $<20$ observations for a subset of individuals). 

Animal movement models have also been developed to account for more mechanistic interactions among individuals (e.g., \citealt{Russell:16}; \citealt{Scharf:15}) and, while we did not address those specifically, the approach we presented may also be beneficial in those settings.  Furthermore, Bayesian animal movement models have been fit using integrated nested Laplace approximation (INLA; \citealt{Rue:09}; \citealt{Illian:12}; \citealt{Illian:13}; \citealt{Ruiz:12}; \citealt{Jonsen:16}) and one could use INLA to fit the hierarchical point process model in our first example.  However, the two-stage MCMC approach presented herein allows for:  Inference on joint relationships among model parameters, easy parallelization in the first stage, and the ability to use Bayesian multiple imputation techniques, such as in our second example involving the CTDS movement model.  

\section*{Acknowledgements}
Support for this research was provided by NSF 1614392, NSF EEID 1414296, CPW T01304, and NOAA AKC188000.  The authors thank The International Environmetrics Society and the editors of Environmetrics for their support and assistance with this work.  The authors also thank Walt Piegorsch, Mindy Rice, Devin Johnson, Peter Craigmile, Erin Peterson, Ron Smith, and the other organizers of the TIES 2016 annual meeting.  Any use of trade, firm, or product names is for descriptive purposes only and does not imply endorsement by the U.S. Government.

\bibliographystyle{wb_env}
\bibliography{animalmovement}

\appendix
\section*{Appendix A: The Imputation Distribution}
\cite{Buderman:16} developed a phenomenological statistical model for estimating an individual's underlying continuous-time path based on Argos telemetry data and a semiparametric regression using temporal basis functions.  We used this model to precalculate an imputation distribution for the true path.  For the $j$th individual, the FMM developed by \cite{Buderman:16} is
\begin{align}
  \mathbf{s}_{ij} &\sim 
  \begin{cases}
    \text{N}(\boldsymbol\mu_j(t_i),\boldsymbol\Sigma_i) &\mbox{ with prob. } p  \\
    \text{N}(\boldsymbol\mu_j(t_i),\mathbf{H}\boldsymbol\Sigma_i\mathbf{H}') &\mbox{ with prob. } 1-p  
  \end{cases} \; , \\ 
  \boldsymbol\mu_j(t_i) &= \mathbf{W}_j(t_i)\boldsymbol\alpha \;, \\ 
  \boldsymbol\alpha &\sim \text{N}(\mathbf{0},\boldsymbol\Sigma_\alpha) \;, 
\end{align}
\noindent where, $\mathbf{s}_{ij}$ represent the $i$th telemetry observation, $\boldsymbol\mu_j(t_i)$ is the true individual position at time $t_i$, $\boldsymbol\Sigma_i$ is an error covariance matrix on the first axis, and $\mathbf{H}\boldsymbol\Sigma_i\mathbf{H}'$ is the error covariance matrix on a rotated axis ($\mathbf{H}$ is a rotation matrix).  The probability $p$ allows the telemetry data to arise from a bivariate Gaussian mixture that captures the X-shaped error pattern inherent to Argos data.  The matrix $\mathbf{W}_j(t_i)$ contains basis vectors (i.e., b-spline basis vectors) at time $t_i$ for individual $j$, and $\boldsymbol\alpha$ is a set of regression coefficients corresponding to the temporal basis functions.  \cite{Buderman:16} set $\boldsymbol\Sigma_\alpha\equiv \text{Diag}(\boldsymbol\sigma^2_\alpha)$ and tuned $\boldsymbol\sigma^2_\alpha$ to induce regularization in the model and improve predictive ability (i.e., ridge regression).         
 
\begin{figure}[htp]
  \centering
  \includegraphics[width=5in, angle=0]{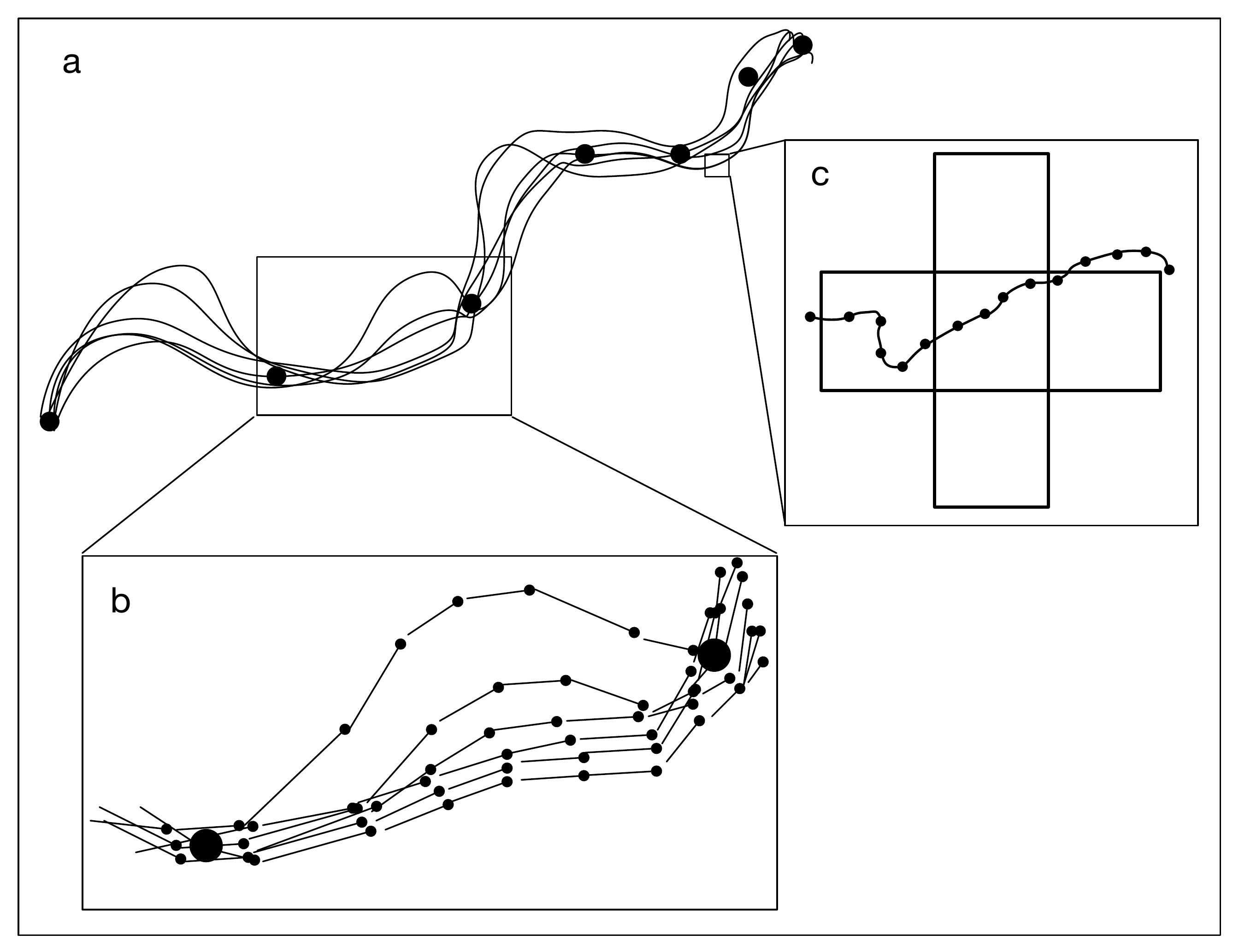}
  \caption{a.) The telemetry data (large points) and imputation distribution (lines) for the individual's path $\{\boldsymbol\mu^k_j(t), \forall k,j,t\}$ using the posterior predictive distribution of the FMM (\citealt{Buderman:16}).  b.) A close up view of the imputation distribution showing the temporal discretization of the imputed path realizations.  c.) A close up view of a single imputed path realization crossing through the first-order neighborhood of the center grid cell.}
  \label{fig:CTDS_FMM}
\end{figure}

The imputed path distribution is obtained by sampling from the posterior predictive distribution of $[\boldsymbol\mu_j(t)|\{\mathbf{s}_{ij}, \forall i,j\}]$ for a large, but finite, set of times $t\in {\cal T}$ to obtain posterior realizations $\boldsymbol\mu^k_j(t)$ for $k=1,\ldots,K$ MCMC iterations.  Figure~\ref{fig:CTDS_FMM}a shows an example set of path realizations (lines) that could result from fitting the FMM from \cite{Buderman:16} to telemetry data (points).  
Figure~\ref{fig:CTDS_FMM}b shows a zoomed in section of the path realizations that highlight the temporal discretization.  At a finer spatial resolution, we can see that the path realizations cross through an example grid cell and its associated neighborhood (Figure~\ref{fig:CTDS_FMM}c).  This idea is critical for processing the path realizations for use with the CTDS model.   

\section*{Appendix B: CTDS Model}
\begin{figure}[htp]
  \centering
  \includegraphics[width=5in, angle=0]{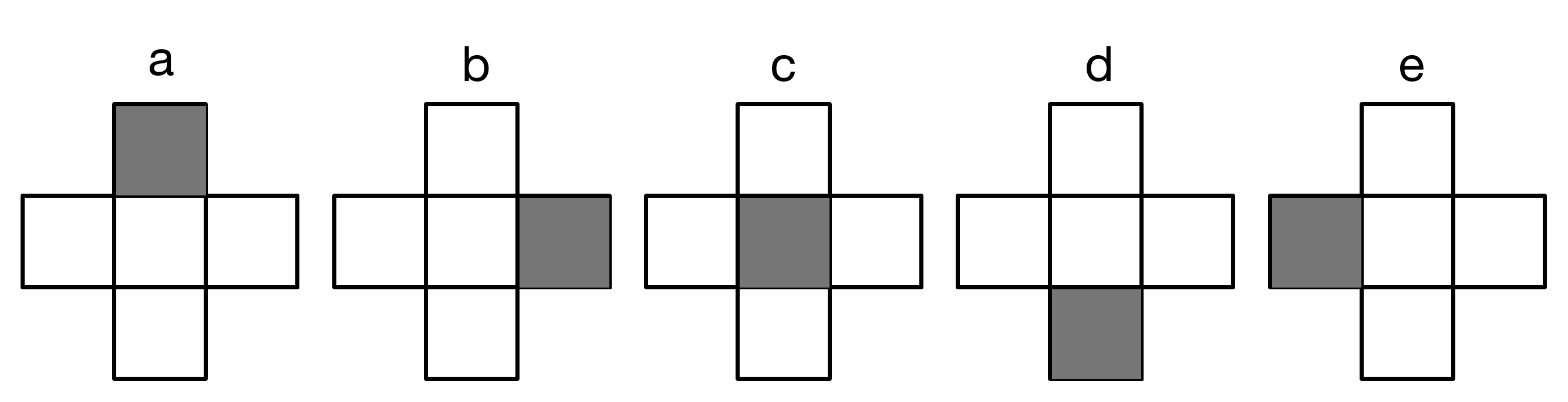}
  \caption{Discrete set of possible transitions at any time $t$, used  to create the multinomial vector $\mathbf{y}(t)$, based on the function $h(\{\boldsymbol\mu_j(t),\forall t\})$.  a.) move up: $\mathbf{y}(t)=(1,0,0,0,0)'$, b.) move right: $\mathbf{y}(t)=(0,1,0,0,0)'$, c.) stay: $\mathbf{y}(t)=(0,0,1,0,0)'$, d.) move down: $\mathbf{y}(t)=(0,0,0,1,0)'$, e.) move left: $\mathbf{y}(t)=(0,0,0,0,1)'$.} 
  \label{fig:moves}
\end{figure}

For each individual $j$ in the original CTDS model, each segment (between points) in Figure~\ref{fig:CTDS_FMM}c served as a multinomial data vector $\mathbf{y}_{ij} \equiv (y_{1i},y_{2i},y_{3i},y_{4i},y_{5i})'_j$ where $\mathbf{y}_{ij} \sim \text{MN}(1,\mathbf{p}_{ij})$ (\citealt{Hooten:10a}).  The multinomial vectors were constructed using the function $\mathbf{y}_{ij}=h(\{\boldsymbol\mu_j(t),\forall t\})$ based on the imputed path realizations by coding a transition as either a stay or a move in a certain direction according to the schematic in Figure~\ref{fig:moves}.  

\cite{Hanks:15a} reparameterized the multinomial imputation data using sufficient statistics.  They denoted residence time as $\tau_{lj}$ (approximated by $\Delta t$ times the number of consecutive stays in the current grid cell, Figure~\ref{fig:CTDS_FMM}c) for $l=1,\ldots,L$ ``stay--move'' pairs and then defined the probability of staying in the current grid cell for time $\tau_{lj}$ as $p^{\tau_{lj}/\Delta t}_{3ij}=(1-p_{lj,\text{move}})^{\tau_{lj}/\Delta t}$, where $p_{lj,\text{move}}$ is the probability of moving.  \cite{Hanks:15a} let $p_{lj,\text{move}}=\Delta t\cdot \lambda_{lj,\text{move}}$ and $\Delta t \rightarrow 0$ yielding     
\begin{equation}
  \lim_{\Delta t \rightarrow 0} (1-p_{j,\text{move}})^{\tau_{lj}/\Delta t} = e^{-\tau_{lj} \lambda_{lj,\text{move}}}  \; ,
\end{equation}
\noindent which, implies that $\tau_{lj} \sim \text{Exp}(\lambda_{lj,\text{move}})$.  

Similarly, \cite{Hanks:15a} showed that the movement probability to neighboring grid cell $c$ is $p_{clj}/p_{lj,\text{move}}=\lambda_{clj}/ \lambda_{lj,\text{move}}$.  Thus, combing the residence probability model with the movement probability yields a likelihood for the sufficient statistic $(\tau_{lj},y_{1lj},y_{2lj},y_{4lj},y_{5lj})'$ equal to $\prod_{l=1}^L \prod_{c\neq 3}\lambda_{clj}\exp(-\tau_{lj}\lambda_{clj})$.  The likelihood for this reparameterized CTDS model coincides with a Poisson where $\lambda_{clj}$ is the movement rate to neighboring cell $c$ and $\tau_{lj}$ is an offset.  Thus, any software capable of fitting a Poisson generalized linear model with an offset can fit the CTDS model if the true path is observed at a fine enough temporal resolution.     

\cite{Hanks:15a} used a multiple imputation approach to account for the uncertainty in the path distribution based on (\ref{eq:mult_imp}).  The movement rates can then be linked to the environmental covariates by a log-linear link $\lambda_{clj}=\mathbf{x}'_{clj}\boldsymbol\beta_j$, where the covariates $\mathbf{x}'_{clj}$ can be specified in several meaningful ways to capture either differential movement rates (i.e., motility) or gradient-based directional bias in movement relative to environmental covariates (see \citealt{Hanks:15a} for details).  The reparameterized CTDS model of \cite{Hanks:15a} is much more computationally efficient than that of \cite{Hooten:10a} because the dimensionality of the data $4L$ depends on the grid cell size instead of the temporal discretization of the path.      

\label{lastpage}

\end{document}